\documentstyle[amstex,amssymb,aps,pra,psfig]{revtex}

\def\nbOne{{\mathchoice {\mathrm{1\mskip-4mu l}} {\mathrm{ 1\mskip-4mu l}}
{\mathrm{ 1\mskip-4.5mu l}} {\mathrm{ 1\mskip-5mu l}}}}

\begin{document}

\title{Scaled Spectroscopy of ${\bf {}^1S^e}$ and ${\bf {}^1P^o}$ 
Highly Excited States of Helium}

\author{B. Gr\'emaud\cite{LKB} and P. Gaspard}
\address{Service de Chimie Physique and Center for Nonlinear Phenomena
and Complex System, \\
Universit\'e Libre de Bruxelles,
Campus Plaine, Code Postal 231, Blvd du Triomphe,
B-1050 Brussels, Belgium.}

\date{\today}
\maketitle
\begin{abstract}

	In this paper, we examine the properties of the ${}^1{\bf S}^e$ and 
${}^1{\bf P}^o$ states of helium,
combining perimetric coordinates and complex rotation methods. We compute
Fourier transforms of quantities of physical interest, among them the average
of the operator $\cos\theta_{12}$, which measures the correlation between the
two electrons. Graphs obtained for both ${}^1{\bf S}^e$ and 
${}^1{\bf P}^o$ states show
peaks at action of classical periodic orbits, either ``frozen planet" orbit or
asymmetric stretch orbits. This observation legitimates the semiclassical
quantization of helium with those orbits only, not just for ${\bf S}$ states 
but also
for ${\bf P}$ states, which is a new result. To emphasize the similarity
between the ${\bf S}$ and ${\bf P}$ states we show wavefunctions of
${}^1{\bf P}^o$ states, presenting the  same 
structure as ${}^1{\bf S}^e$ states, namely the ``frozen planet"
and asymmetric stretch configurations.

\end{abstract}

\pacs{PACS number(s): }


\section{INTRODUCTION}

	The helium atom is one of the prototypes of atomic
systems showing chaotic behaviour at the classical level. Even if the
non-integrability of the three-body problem has been known 
for a long time, major steps
have been realized only during the last ten years, giving rise to a
regain of activity in this field~\cite{Richter90,Richter902,Ezra91,Blumel91,%
Richter91,Richter92,Gaspard93,Tanner95,Burgers95,Burgers95}. 
From a theoretical point of view, two reasons have
contributed to recent advances. One of the reasons has been the development of
powerful methods for the numerical resolution of the quantum problem along with
the advent of modern computers. The other reason has been the emergence of 
an understanding of the quantum mechanics of classically chaotic systems,
especially, by the developments of new semiclassical methods, which link the
two different worlds \cite{Gutzwiller90}. In addition, from the experimental point
of view, the most recent technologies have enabled very high resolution
spectroscopy of the helium atom, reaching energy domains where  different
Rydberg series strongly overlap giving rise to chaotic behaviour
\cite{Domke91,Domke92,Domke95,Domke96}. 
As a consequence of all these studies, the helium atom is
now quite well  understood both from the quantum and the semiclassical point of
view. 

	Still, the knowledge on this subject is poor compared to that of other
atomic systems, like the hydrogen atom in a magnetic field or in a microwave
field. For example, there are no systematic studies of the 
classical phase space, only a few types of periodic orbits are known.
Surprisingly, these orbits suffice for the semiclassically quantization of the
helium atom or of the negative hydrogen ion with quite a good agreement with
{\it ab initio} quantum 
calculations~\cite{Wintgen92,Muller92,Gaspard93,Richter91}. One
purpose of this paper is the validation of these results, by proceeding the
other way round~: extracting the classical information  from the quantum
properties. Although, this kind of study has already been  conducted for
a long time on many other chaotic systems, it does not exist for the
helium atom. There are many reasons to explain this lack, among them the fundamental
one is probably the high complexity of the numerical resolution of the full
quantum problem ( all degrees of freedom and no approximations). The same is
also true for the classical approach~: a very large phase space (eight
dimensional) and a dynamics which is never fully chaotic.

	Through intensive numerics, we 
have been able to overcome these difficulties and
to perform a semiclassical analysis of the helium spectrum, using the scaled
Fourier transform method. Thus, we have been able to recover the actions of the
collinear periodic orbits (both $eZe$ and $Zee$ configurations) 
from the $L=0$ states. As the experimentally accessible states are $L=1$
states, we applied the same method to them. Similar results have then been
obtained, in agreement with the usual treatment of non-zero angular
momentum states with $L=0$ periodic orbits. The similarity between the $L=0$
and $L=1$ states is emphasized by plotting wavefunctions of $L=1$ states,
showing the same kind of structures, namely frozen-planet configuration or
localization around the asymmetric stretch orbit.

	The paper is organized as follows~: In Sec. \ref{quantum}
we summarize the generalities about
the quantum helium and its numerical resolution. In Sec. 
\ref{classical}, we describe the properties of the classical system. 
In Sec. \ref{results} we present the numerical scaled
spectroscopy of  ${}^1{\bf S}^e$ and ${}^1{\bf P}^o$ states, 
with wavefunctions of 
${}^1{\bf P}^o$ states. These results and the conclusions are discussed in
Sec. \ref{conclusion}.

\section{HELIUM: EXACT QUANTUM ANALYSIS}
\label{quantum}

\subsection{\label{gen} Generalities}

\begin{figure}
\centerline{\psfig{figure=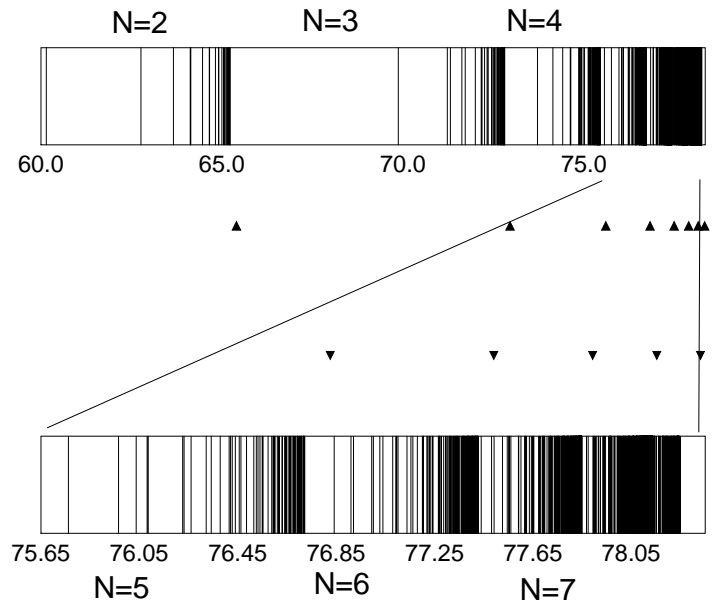,height=8cm,angle=-90}}
\vspace{0.5cm}
\caption{\label{series} Spectrum of the helium atom (${}^1{\bf P}^o$ states)
as it is obtained from numerical diagonalization. Energies are given in
electron-Volt above the ground state. Only the resonances are
shown ($8\geq N\geq 2$).
It is made of Rydberg series converging to simple ionization thresholds
($\blacktriangle$ and $\blacktriangledown$), which are part
of the Rydberg series of the He$^+$ levels converging towards the double
ionization limit ($E=79.003$ eV). (Because of the numerical precision, the 
calculated Rydberg series are truncated, so that they appear to converge
to effective thresholds lower in energy). Above the $N=3$ threshold, the
series overlap. }
\end{figure}

	 The schematic structure 
based on the independent electrons picture is rather simple~: the
first electron gives rise to a Rydberg series (the He$^+$ levels) of simple
ionization thresholds.  The second electron creates further Rydberg series
which converge to each of the simple ionization thresholds (see
Fig.~\ref{series}).  The limit of the series of the simple ionization thresholds
is the double ionization threshold. Above the first simple ionization threshold,
all states are doubly excited states and because of the electron-electron
interaction they become resonances~:  one electron can ionize as the other falls
on a lower state. The first series remains discrete. The different series are
labelled by the quantum number $N$, which is the principal quantum number of the
He$^+$ hydrogenic levels.

	After the first experimental observations of the doubly excited 
states~\cite{Madden63},
it has become clear that this picture was unable to give any
sensible quantum numbers. Many different systems have been proposed to
label the states~\cite{Sinanoglu73,Herrick75,Sinanoglu75,Herrick77,Herrick80,%
Herrick83,Feagin86,Feagin88,Rost88,Kellman94}. 
Here we only present  the most frequently used, proposed by
Herrick and Sinanoglu. Apart from the two obvious quantum number $(L,M)$ 
associated
with the total angular momentum and its projection along an arbitrary fixed
axis, they introduced three other quantum numbers $(N,K,T)$. $N$ is the
principal quantum number of the inner electron (the threshold label). $T$ is
the projection (actually, the absolute value) 
of the total angular momentum onto the inter-electronic axis. 
$K$ is related to the projection of the
difference of the two Runge-Lenz vectors on the inter-electronic axis. For
very high excited states the quantum number $K$ measures the angular correlation
between the two electrons. Indeed, the expectation value of the operator
$\cos\theta_{12}$ in the $|N,K,T\rangle$ states is $-K/N$ for $N$ large.
Finally, an additional quantum number is added to label the behaviour of the
states under the exchange of the two electrons.

\subsection{Perimetric coordinates}

	Different methods are used to solve numerically the non-relativistic
three-body Coulomb
problem~\cite{Ho86,Watanabe86,Dmitrieva86,Rost91,Briggs91,Ho93,Ostrovsky93,%
Muller94}. Among these, the most efficient one is probably the combination of 
perimetric coordinates and complex rotation. This method is exact without
any approximation and takes into account all the degrees of freedom. A complete
review  of it will be given in~\cite{perimetric}. Not only does it provide the
energies and the widths of the resonances, but it also allows us to compute
any quantities of physical interest like oscillator strengths, wavefunctions...
A short compilation is noted as follows~:

	The non-relativistic Hamiltonian, with infinite nucleus mass can be
written in atomic units ($\hbar=m_{e^-}=4\pi\epsilon_0=e^2=1$) as~:
\begin{equation}
\label{hamil}
 H=\frac 12 ({\bf P}^2_1+{\bf P}^2_2) -\frac 1{r_1} -\frac 1{r_2} 
+\frac 1Z \frac 1{r_{12}} 
\end{equation}

\noindent where the scaling $\bbox{r}_i \rightarrow Z\bbox{r}_i$ has been
carried out, so that for $Z$ large, we are left with two independent hydrogen
atoms perturbated by the term $1/(r_{12}Z)$. Because of this scaling, the
eigenenergies of the Hamiltonian~\eqref{hamil} must be mutiplied by the
factor $Z^2$ to get the eigenenergies of the original system. For 
instance the ground state energy of helium ($Z=2$) is $-0.725931094$ 
units in our energy scale (instead of $-2.903724377$ a.u.).

\begin{figure}
\centerline{\psfig{figure=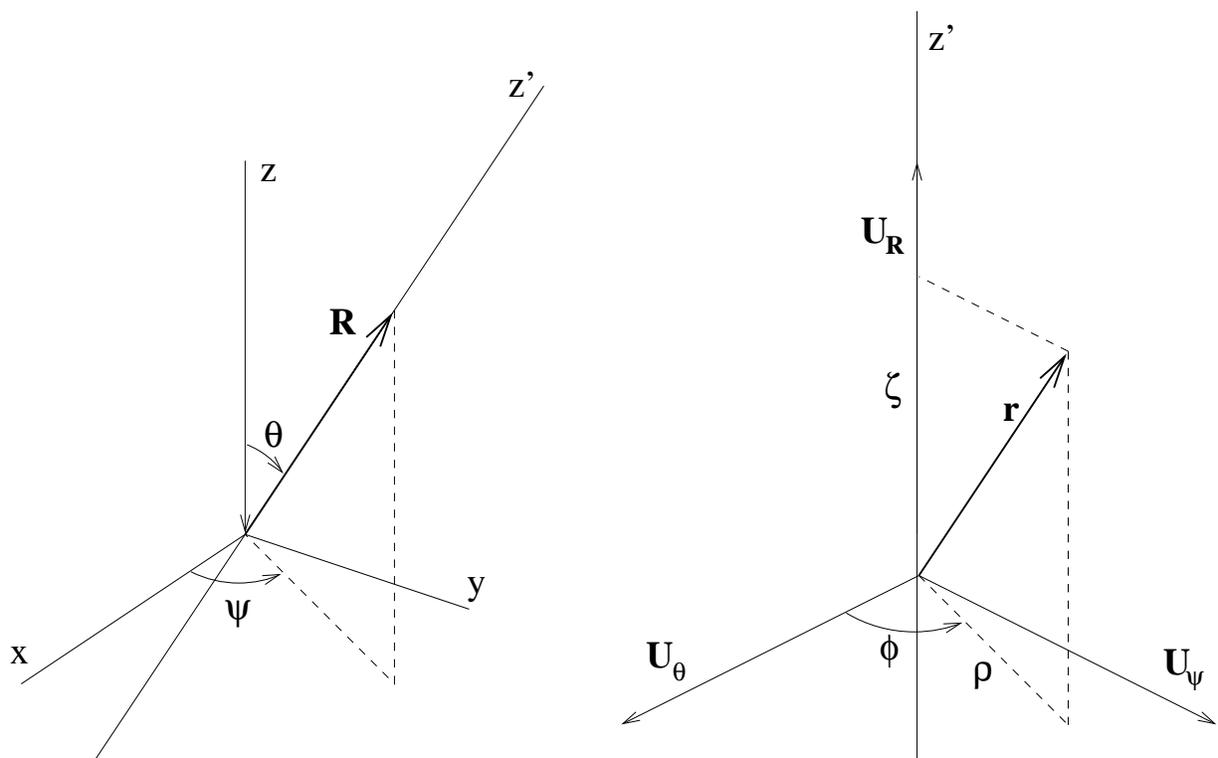,height=10cm,angle=-90}}
\caption{\label{coordonnees} Definition of the molecular-like system
of coordinates. ${\bf R}$ is the inter-electronic vector and ${\bf r}$ 
is the vector from the origin to the centre of mass of the two electrons. 
${\bf R}$ is defined by the usual spherical coordinate $(R,\theta,\Psi)$  
and $(\rho,\zeta,\phi)$ are the cylindrical coordinates of ${\bf r}$
relatively to the body-fixed axis $z'$ defined by ${\bf R}$}
\end{figure}

	The rotational invariance of $H$ is used to separate the
angular dependency of the wavefunction from the relevant dynamical
variables, namely the three inter-particule distances. For a given pair of
good quantum numbers ($L,M$) (total angular momentum and its
projection onto the $z$-axis of the fixed laboratory frame), the wavefunctions
have the following expression~:

\begin{equation}
\Psi_{L\,M}=\sum_{T=-L}^L {\cal D}^{L^{\displaystyle *}}_{MT}
(\psi,\theta,\phi) \Phi_{T}^{(LM)}(R,\rho,\zeta)
\end{equation}

\noindent where $(\psi,\theta,\phi)$ are Euler angles defined by the 
transformation of the original fixed frame to a molecular-like frame 
moving with the inter-electronic axis (see figure~\ref{coordonnees}). 
The functions ${\cal D}^{L^{\displaystyle *}}_{MT}(\psi,\theta,\phi)$
are the matrix elements of the rotation operator in the $L$ representation. 
They are eigenfunctions of the operators $\lbrace{\bf L}^2,L_z,L_z'\rbrace$
for the eigenvalues $\lbrace L(L+1),M,T\rbrace$, where $L_z'$ is the projection
of the angular momentum along the inter-electronic axis.
The three remaining variables $(R,\rho,\zeta)$ are actually 
combinations of $r_1$, $r_2$ and $r_{12}$. All the information
about the dynamics is thus contained in the various functions $\Phi_{T}$, which
are coupled by Coriolis-like terms. The Hamiltonian $H$ is then prediagonalized
and we are left with effective Hamiltonians $H_{J}$ in which all the angular
variables are stripped out. Eventually, these Hamiltonians
become matrix Hamiltonians. Finally, the two discrete symmetries (parity and
exchange of the two electrons) add relations between the functions $\Phi_T$. For
example, the ${\bf S}$ states have only one component $\Phi(=\Phi_0)$, which
can be chosen symmetric (${}^1{\bf S}^e$ states) or antisymmetric 
(${}^3{\bf S}^e$ states). On the other hand, the ${\bf P}$ states are
described by three components $\lbrace\Phi_T\rbrace_{T=0,\pm 1}$.  The
${\bf P}$ states separate between odd and even states and
finally, for odd states, the {\it a priori} three independent $\Phi_T$'s
reduce to only one component, but with no more symmetry with respect to the
exchange of the two electrons.

	Although the molecular-like coordinates $(R,\rho,\zeta)$ are useful 
to exploit the rotational invariance, they are not suitable for the numerical
implementation. To overcome this difficulty, the well-known perimetric
coordinates are introduced~:

\begin{equation}
\left\{ \begin{array}{cr}
x= & r_1+r_2-r_{12} \\
y= & r_1-r_2+r_{12} \\
z= & -r_1+r_2+r_{12} 
\end{array} \right. 
\end{equation}

\noindent Their domains of variation range from $0$ to
$+\infty$ and are independent. (In contrast, we would have the condition
$|r_1-r_2|\leq r_{12}\leq r_1+r_2$ between $r_1$, $r_2$ and $r_{12}$).
	 
	The effective Hamiltonians are then expanded in the product of three
``Sturmian" basis, one for each perimetric coordinates. The basis functions  have
the following expression~:

\begin{equation}
\label{basisfns}
\phi_n^{\alpha_u}(u)=\sqrt{\alpha_u}L_n(\alpha_u u)e^{-\frac{\alpha_u}2 u}
\end{equation}

\noindent	where $u$ is one of the perimetric coordinates $(x,y,z)$,
$n$ a positive integer, $\alpha_u$ a real parameter (the scaling or dilatation
parameter) and $L_n$ the $n^{\mathrm{th}}$ Laguerre polynomial. 
These functions are  associated with the representation of the group $SO(2,1)$
formed by the following operators~:

\begin{equation}
\left\{		
\begin{split}
S_3^{\alpha_u} &= -\frac 1{\alpha_u}\left[u\frac{\partial^2}{\partial u^2}
+\frac{\partial}{\partial u}\right]+\alpha_u \frac u4 \\
S_1^{\alpha_u} &= \phantom{-}\frac 1{\alpha_u}\left[u\frac{\partial^2}
{\partial u^2}+\frac{\partial}{\partial u}\right]+\alpha_u \frac u4 \\
S_2^{\alpha_u} &=\imath\left[u\frac{\partial}{\partial u} +\frac 12\right]
\end{split} \right.
\end{equation}

\noindent Thanks to their group properties, these operators have selection rules
in the Sturmian basis and the few non-zero matrix elements are analytically
known. As the effective Hamiltonians are polynomial functions
 of the nine operators
$S_i^u$ ($i=1,2,3$ and $u=x,y,z$), they also have selection rules. For
${\bf S}$ (resp. ${\bf P}^o$) states there are only $57$ (resp. $215$)
non-zero matrix elements whose expressions are also given by
polynomials. The representations of the
$H_{J}$ in the Sturmian basis are then infinite sparse banded matrices,
which can be efficiently diagonalized (after truncation) using the Lanczos
algorithm. This iterative method is extremely efficient 
for computing few eigenvalues of a 
very large banded matrix (typically $100$ for a $10000\times 10000$ matrix
with bandsize equal to $1000$) in a very short time ($\simeq 10$ minutes 
CPU on a usual workstation).

\subsection{Complex rotation}

	As stated in Section~\ref{gen}, above the first threshold of
simple ionization, all the states are resonances. The complex rotation
method is appropriate to compute directly the properties of these resonances
(energy, width). Its properties rely on deep mathematical properties
 of the analytic
continuation of the Green function in the complex plane 
\cite{Ho83,Balslev71}. A recent review of its application to atomic physics can 
be found in Ref.~\cite{Buchl94}. 

\begin{figure}
\centerline{\psfig{figure=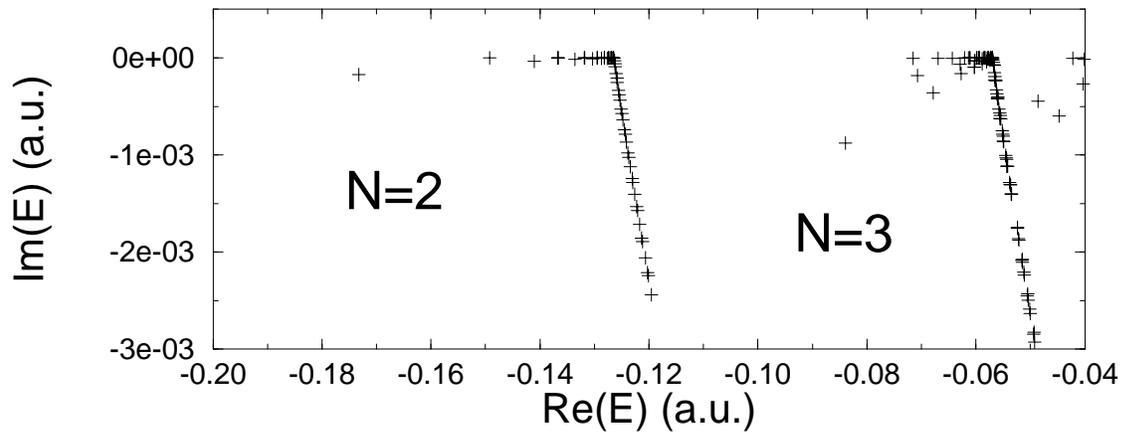,height=14cm,angle=-90}}
\caption{\label{complex} Complex sprectrum of ${}^1{\bf P}^o$ 
states of the helium
atom showing the $N=2$ and $N=3$  Rydberg series. This
spectrum is obtained by numerical diagonalization of the complex Hamiltonian
$H(\theta)$ for $\theta=0.16$. Due to the matrix truncation, the rotated
continua are discretized.}
\end{figure}

	The method is included in our case by making the scaling parameters
$\alpha_u$ complex, more precisely $\alpha_u \rightarrow \alpha_u e^{-\imath 
\theta}$, where $\theta$ is a real parameter (the rotation angle). 
The matrix representations of the Hamiltonians then become complex symmetric, 
but are no more Hermitian. The
fundamental properties of the complex spectra are (see Fig.~\ref{complex})~:

\begin{itemize}

\item The bound states are still on the real axis.

\item The continua are rotated by an angle $2\theta$ on the lower-half complex
plane, around their branching point. 

\item Each complex eigenvalue $E_i$ gives the properties of one resonance, i.e.
the energy is the real part of $E_i$, and the width is two times the negative
of the imaginary part. The complex eigenvalues are independent of $\theta$,
provided that they are not covered by the continua.

\end{itemize}

	Therefore, the rotated Green function $G(\theta)$ associated with
$H(\theta)$ has no longer a branch cut along the real axis and thus directly
gives  the analytic continuation of the original Green function in the complex
lower half-plane. For a negative value of $\theta$, one obtains the analytic
continuation in the upper half-plane. Furthermore, the eigenbasis of $H(\theta)$
satisfies the two properties of closeness and orthogonality (for a non-Hermitian
scalar product), and is thus used to express all the physical quantities
originally written in the eigenbasis of $H$. For example, the one-photon 
photoionization cross-section from the ground state is given by~:

\begin{equation}
\label{crossfor}
\sigma(\omega)=\frac {4\pi\omega}{c}\mathrm{Im}
\sum_i\frac{\langle\overline{E_{i\theta}}|R(\theta)T|g\rangle^2}{E_{i\theta}
-E_g-\hbar\omega} 
\end{equation}

\noindent where $\hbar\omega$ is the photon energy, $E_g$ is the 
ground state energy and $T$ is the dipole operator. The main difference
between the usual expression ($\frac {4\pi\omega}{c}|\langle E|T|g\rangle|^2$)
and formula~\eqref{crossfor} 
is the contribution of all eigenstates $|E_{i\theta}\rangle$ of 
$H(\theta)$ rather 
than only the continuum state $|E\rangle$ of $H$. There are other little
differences due to the non-Hermiticity of $H(\theta)$~: for
instance the complex
conjugate of $\langle E_{i\theta}|$, $\langle\overline{E_{i\theta}}|$ and the
square of the matrix element (not the modulus square) enter the
formula~\eqref{crossfor}.  Also, the presence of the operator $R(\theta)$ is
necessary to backward-rotate $\langle\overline{E_{i\theta}}|$ onto the real
axis. A consequence is that for an uncovered resonance, the matrix element 
$\langle\overline{E_{i\theta}}|R(\theta)T|g\rangle$ is independent of $\theta$,
enlightening the underlying analytic properties of the Green function. On the
contrary, the matrix element $\langle\overline{E_{i\theta}}|T|g\rangle$ is
$\theta$-dependent and has thus no clear physical meaning, although some
interpretations have recently been extracted~\cite{Burgers96}.
In principle, for a given energy $E$, only the total sum has a physical meaning,
but for a well-isolated resonance $E_{i\theta}$ (whose
width is much smaller than the distance in energy to the other ones) and for
$E\simeq\mathrm{Re}(E_{i\theta})$, the main contribution comes from this
resonance only with a Fano profile shape, whose $q$-parameter is given by~:

\begin{equation}
q=-\frac{\mathrm{Re}\langle\overline{E_{i\theta}}|R(\theta)T|g\rangle}
{\mathrm{Im}\langle\overline{E_{i\theta}}|R(\theta)T|g\rangle}
\end{equation}

\noindent Finally, in the general case, thanks to the energy denominators, 
only the resonances with energies close to the energy $E=E_g+\hbar\omega$ will
contribute to the sum, which  allows a numerical efficiency. 

The complex rotation method can also be used to compute the
probability density of a continuum state of energy $E$, through the following
formula~:

\begin{equation}
|\Psi_E({\bf {\bf r}})|^2=\langle {\bf r}|E\rangle^2=
\frac 1{\pi}\sum_i\frac{\langle{\bf r}|R(-\theta)|E_{i\theta}
\rangle^2}{E_{i\theta}-E}
\end{equation}

\noindent Again, for a well-isolated resonance, the main contribution of the
preceding sum [at the energy $E=\mathrm{Re}(E_{i\theta})$] comes from this
resonance, more precisely~:

\begin{equation}
|\Psi_{E\sim E_{i\theta}}({\bf r})|^2\sim\frac 1{\pi|\mathrm{Im}
(E_{i\theta})|}\mathrm{Re}\lbrack\langle{{\bf r}|R(-\theta)|E_{i\theta}
\rangle^2}\rbrack
\end{equation}

\noindent which may thus be interpreted as the probability density of one
resonance. Again, the backward rotation is essential to obtain sensible
results and to provide physical interpretation.

\section{HELIUM: SEMI-CLASSICAL ANALYSIS}
\label{classical}

\subsection{Periodic orbits}

	The main difficulty of the classical dynamics of the helium atom
is the dimension of the phase space. Even for $L=0$ trajectories, the
number of degrees of freedom is three, so that usual techniques like Poincar\'e
surfaces of section are of no help.  For non-zero angular momentum, there is
another degree of freedom, which further increases the complexity.  A systematic
study  of the classical dynamics is still needed. Another singularity is that the
motion is never bounded, even for negative energies. Thus classically, one
electron can ionize at any energy, in contrast to the quantum problem for which
exact bound states do exist.  All this explains why so few periodic orbits are
known. 

Two classes of periodic orbits have already been widely explored~: the 
collinear orbits and orbits on the Wannier ridge
\cite{Richter902,Richter93,Muller92}. The latter is of relatively little
interest with respect to the quantum problem  (at least for $N\leq 10$, see
Ref.~\cite{Muller92}).  On the contrary, the first class has been
extensively used for the semiclassical quantization.  For these orbits, the two
electrons and the nucleus stay on the same axis at all times. They are of two
types according to whether the electrons are on the same side of the nucleus or
on opposite sides. 

When the electrons are on opposite sides, the classical orbits
are asymmetric stretches made of successive rebounds of one or the other
electron on the nucleus. This simple scheme is used to encode the periodic
orbits and, thus, to semiclassically quantize the helium atom through
cycle-expansion techniques \cite{Wintgen92,Tanner95}. 
These orbits are unstable with
respect to collinear perturbation and stable for perturbation in the other
directions. 

	When the electrons are on the same side of the nucleus, there exists
one periodic orbit, the now well-known frozen-planet configuration, which is
surprisingly stable with respect to perturbations in any directions. The motion
of the two electrons is extremely different from the asymmetric stretches~:  the
outer electron is dynamically frozen at a large distance from the nucleus, while
the inner electron oscillates in the electric field produced by the outer one. 
The torus quantization of this orbit has been fruitful to explain the long-lived
resonances of the helium atom \cite{Wintgen92}.

	In Table~\ref{po}, we give the reduced actions ($S_p/2\pi$) and others
properties of the various periodic orbits.

\subsection{Scaling law and Fourier transform}

	The connection between the quantum and the classical worlds is made,
in the case of a classically chaotic systems,
by the Gutzwiller trace formula, which expresses the oscillating part of the
density of states as an (infinite) sum over all unstable periodic orbits 
and their repetitions involving classical quantities 
only~\cite{Gutzwiller90,Gaspard95}.
This formula is a particular case of the following more general expression~:

\begin{equation}
\label{trace1}
\mathrm{tr}\left.\frac {\hat{A}}{E-\hat{H}}\right|_{\mathrm{osc}}=
\frac 1{\imath\hbar}\sum_{p\,r}\left(\oint_p A_W\,dt\right)
\frac {\exp{\left[\frac {i}{\hbar}rS_p(E)-i\frac{\pi}2 r\mu_p\right]}}
{|\mathrm{det}({\bf m}^r_p-\nbOne)|^{1/2}}
\end{equation}

\noindent where $\hat{A}$ is an arbitrary operator ($\cos\theta_{12}$ in our
case) and $A_W$ is its Weyl-Wigner representation. $S_p(E)$ is the reduced action
of the periodic orbit $p$ at the energy $E$, $\mu_p$ is its Maslov index, and
${\bf m}_p$ is the monodromy matrix associated with this orbit.  For 
$\hat{A}=\nbOne$, we recover the usual oscillatory part of the density of
states.

	The scaling law of the Coulombian dynamics gives directly the
dependence of the reduced action with respect to the energy. For
$E<0$ we have~:

\begin{equation}
S_p(E)=\frac 1{\sqrt{-E}} S_p(-1)=\frac 1{\sqrt{-E}}S_p
\end{equation}

\noindent Thus, any periodic orbit at a negative energy $E$ is mapped onto 
the  same orbit at energy equal to $-1$. The right-hand member of
equation~\eqref{trace1} becomes~:

\begin{equation}
\frac 1{(-E)^{3/2}}\sum_{r\,p} {\cal A}(r,p)\exp{\left[ 
\frac {i}{\hbar} \frac 1{\sqrt{-E}} r S_p(-1)\right]}
\end{equation}

\noindent In the preceding expression, ${\cal A}$ denotes an amplitude factor
independent of the energy. We also recover the result that the semiclassical
limit  is reached when $E$ goes to zero, i.e., for very highly excited
states.

	Furthermore, by taking the Fourier transform of $\mathrm{tr} \frac
{\hat{A}}{E-\hat{H}}$, with respect to the variable $1/\sqrt{-E}$, we should
obtain peaks at the scaled reduced actions $S_p$ of the periodic orbits and their
repetitions~: This method is called the scaled spectroscopy and has
already shown its power in many other chaotic systems.

\section{NUMERICAL RESULTS}
\label{results}

\subsection{Numerical implementation}

	For numerical purpose, we have to truncate the Sturmian basis. Since a basis
vector is labelled with the three quantum number $(n_x,n_y,n_z)$, we truncate 
the basis with $n_x+n_y+n_z \leq N_{\mathrm max}$. We also work with
$\alpha=2\beta= 2\gamma$ where $\alpha=\alpha_x$, $\beta=\alpha_y$, and
$\gamma=\alpha_z$ are the three dilatation parameters of the basis functions
\eqref{basisfns}. This choice gives a good asymptotic behaviour of the
wavefunction for large values of both $r_1$ and $r_2$. This choice also
introduces additional selection rules, which therefore increase the sparseness
of the matrices. For the ${}^1{\bf S}^e$ states, we can also  symmetrize the
basis and consider $|n_x,n_y,n_z\rangle + |n_x,n_z,n_y\rangle $.
 We have considered truncations up to $N_{\mathrm max}=58$, which gives a
basis size equal to 18445 and a bandsize equal to 932. On a CRAY J90, the
codes typically take 1000 seconds to run. For the ${}^1{\bf P}^o$ states, we
have also carried out the numerical calculation up to $N_{\mathrm max}=58$ where
the basis size is equal to 35990 and the bandsize is equal to 3056. Here, the
codes take 2000 seconds of CPU time to run on a CRAY C98. The
convergence of the results have been checked with systematic variation of the
basis size, of the complex-rotation angle $\theta$, and of the dilatation
parameter $\vert\alpha_u\vert$.

\subsection{Photoionization cross-section}

For the comparison with the experiments, the photoionization cross-section from
the ground state is a suitable quantity for many different
reasons. The first one is the existence of high quality experimental 
results~\cite{Domke95,Domke96}, especially in the  region where series 
strongly overlap. Furthermore, at the experimental resolution, neither the
QED nor relativistic effects are resolved, so that no correction needs to be
added to our results, in the present stage of experimental accuracy. Finally,
because the cross-section involves the computation of the (complex) oscillator
strengths, it will test not only the good convergence of the complex
eigenenergies, but also of the good implementation of the backward rotation and
the overlaps between eigenvectors computed at different scaling parameters,
which is far from obvious~\cite{perimetric}. Especially, the Fano $q$-parameter
values are strongly dependent on the backward rotation.

\begin{figure}
\centerline{\psfig{figure=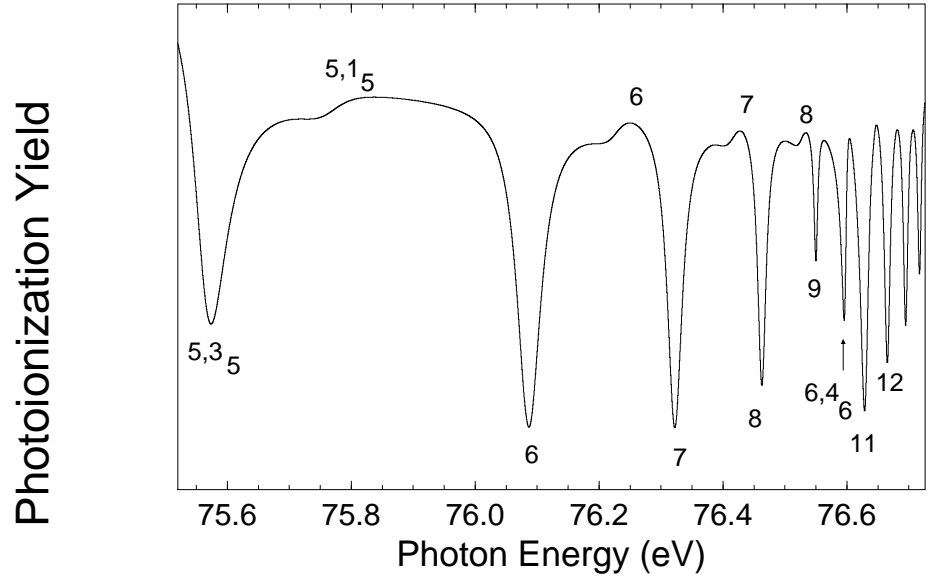,height=12cm,angle=-90}}
\caption{\label{cross} Theoretical photoionization cross-section below the
$N=5$ threshold, convoluted at the experimental resolution of 
reference~\protect\cite{Domke95}. The oscillations due to overlap with the
$6,4_6$ member from the upper series are very well reproduced. The resonances
belonging to the two main series,$5,3_n$ (below) and $5,1_n$ (top), 
are indicated using Lin's classification scheme. The position of the
perturber coming from the $N=6$ series is also indicated.}
\end{figure}

\begin{figure}
\centerline{\psfig{figure=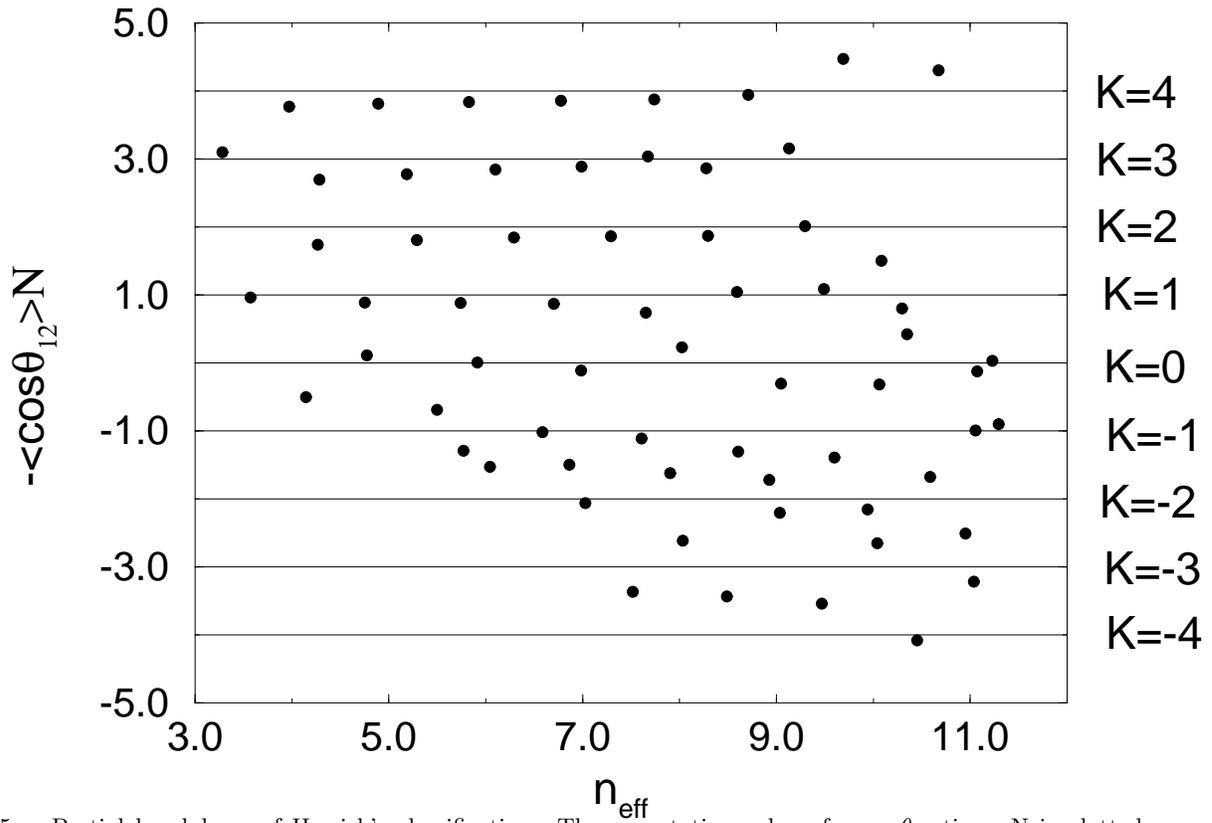,height=12cm,angle=-90}}
\vspace{0.5cm}
\caption{\label{KT} Partial breakdown of Herrick's classification. The
expectation value of $-\cos\theta_{12}$ times N is plotted versus the
effective quantum number $n_{\mathrm{eff}}$. Although series with high
$K$ number are well separated, the others are mixing, especially in the
energy range where interaction with the $6,4_6$ state occurs. Horizontal
lines despict the values predicted by Herrick's theory.}
\end{figure}

	Our results are based on the infinite nucleus-mass Hamiltonian. 
The corrections are taken into account by using the effective double ionization
threshold value and the Rydberg constant value given in
Ref.~\cite{Domke96}.  Furthermore, we have convoluted our ``infinite
precision" cross-section with a Lorentzian of width equal to the experimental
resolution. Fig.~\ref{cross} gives our numerical cross-section below the
$N=5$  
threshold where oscillations due to the interference with the lowest
members of the $N=6$ series appears. The agreement with the figure of
Ref.~\cite{Domke95} is very good. We emphazise that there are no adjustable
parameters. To make a better comparison, we give in Table ~\ref{data} all the
characteristic quantities (energy, reduced linewidth, reduced probability
transition, Fano $q$-parameter,  value of the inter-electronic angle)
we have calculated in this energy range for the two principal series
$5,3_n$ and $5,1_n$. Even if still valid, Herrick's classification starts to
partially break up, as one can see in Fig.~\ref{KT}, where
$-\langle\cos\theta_{12}\rangle N$ is plotted versus the effective quantum
number $n_{\mathrm{eff}}$. This kind of plot has already been observed and
analysed in Ref.~\cite{Burgers95}. Thus, the conclusion of
Ref.~\cite{Burgers95} extends to the ${}^1{\bf P}^o$
states, which is of great interest since the ${}^1{\bf P}^o$ states are
accessible in the current experiments.  A more complete description of the
photoionization cross-section will be given in Ref.~\cite{section}.

\subsection{Scaled spectroscopy}

	As explained in the Introduction, one of our goals is to retrieve the
classical quantities from the quantum ones. Thus, we have applied the
scaled Fourier transform (SFT) techniques to the
trace of the Green function (i.e. the density of states) and also to the
trace of the operator $G(E)\times \cos\theta_{12}$, which characterizes the
electron-electron correlations. Furthermore, since the function 
$\theta\rightarrow\cos\theta$ has zero derivative for $\theta=0$ or
$\theta=\pi$, it enhances the weight of the collinear orbits. The SFT is
convoluted with a Welch window in order to lower the effects of the finite size
of our spectrum.

\begin{figure}
\centerline{\psfig{figure=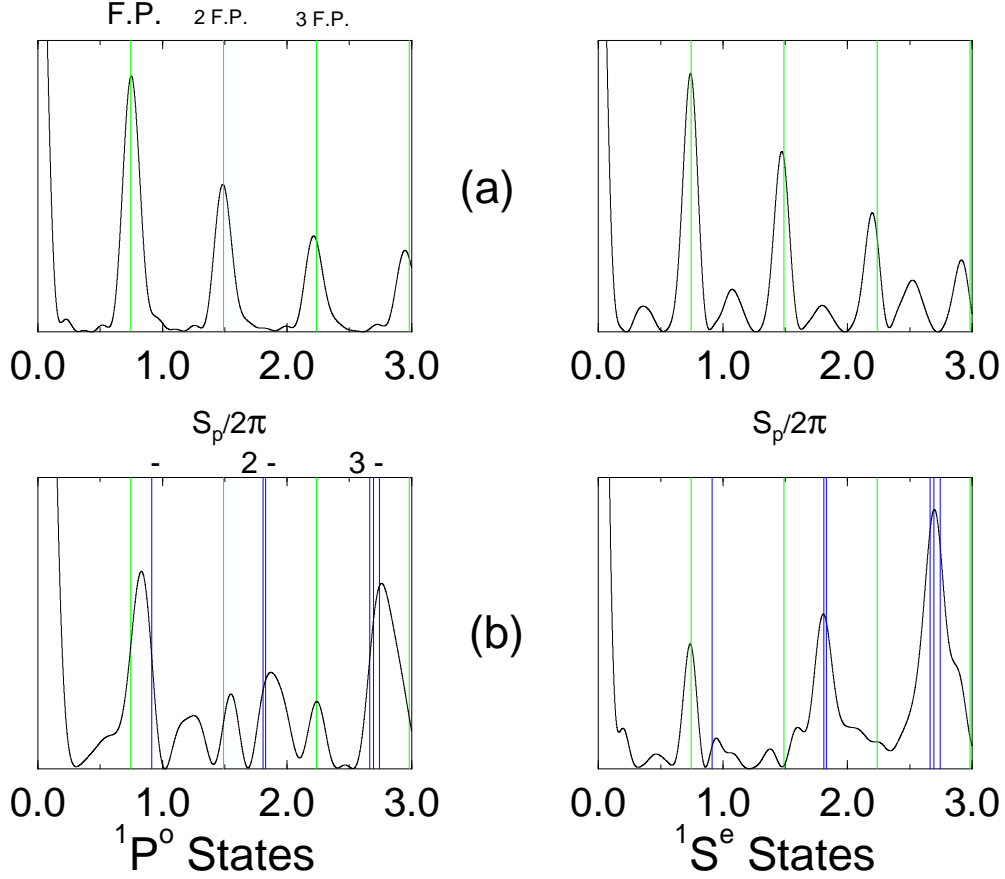,height=12cm,angle=-90}}
\vspace{1cm}
\caption{\label{trace} Scaled spectroscopy of  
${}^1S^e$ states (on the right) and ${}^1P^o$ states 
(on the left). The upper plots (a) depict the SFT of the
Green function $G(E)$. The vertical lines gives the values 
of reduced action of the
frozen-planet orbit ($Zee$ configuration) and its repetitions. Both SFTs
have peaks at those actions. The lower plots (b) depict the SFT of the
operator $\cos\theta_{12} G(E)$. The weight of the $Zee$ orbits have been
lowered, while peaks appear at the reduced actions of 
three first $eZe$ configuration orbits 
and their repetitions (given by the solid vertical lines,
the dotted vertical lines give the repetitions of the action of the $Zee$
orbit). $-$ corresponds to the asymetric strech, $2-$ (resp. $3-$) to its
second (resp. third) repetition, which is very close to the $+-$ 
(resp. $++-$ and $+--$) orbit.} 
\end{figure}

	 Fig.~\ref{trace} depicts the different
results obtained for the ${}^1{\bf S}^e$ and ${}^1{\bf P}^o$ states. 
At the top, the SFT of the
density only is shown, the ${}^1{\bf S}^e$ case being on the right-hand
side.  The vertical lines indicate the scaled classical action of the
frozen-planet periodic orbit ($Zee$ configuration)  and their repetitions. For
both the ${}^1{\bf S}^e$ states and the  ${}^1{\bf P}^o$ states,
the SFT clearly has peaks at those classical actions. The fact that the
$Zee$ orbit is the only one visible is probably related to its stability in all
directions. At the bottom of Fig.~\ref{trace}, we plotted the SFT of
$G(E)\times\cos\theta_{12}$. The peaks at the  $Zee$ orbit have been
lowered and some have disappeared. For the
${}^1{\bf S}^e$ states, we see new dominant peaks at the classical actions of
the three first asymmetric stretch orbits ($eZe$ configuration). For the
${}^1{\bf P}^o$ states, we also see new dominant peaks we can associate
with the $eZe$ asymmetric stretch orbits.  However, the agreement is not so
good because the peaks are slightly shifted although the $eZe$ orbits are still
emerging from the quantum data. In all these plots, the shifts tends to decrease,
when increasing the number of eigenvalues used to compute the traces. 
	
	The fact that peaks at $L=0$ orbits are obtained for $L\ne 0$ might
be surprising, but this can be explained from 
a semi-classical point of view. For
a given value of the quantum total angular momentum, we obtain an 
effective Hamiltonian acting only on the radial coordinates, in which
centrifugal potentials (inverse square) appear. Taking the classical
equivalent of this Hamiltonian and using the Coulomb scaling law, we
directly obtain that these potentials vanish as $|E|\rightarrow 0$. 
This means that their
contributions to the classical dynamics become negligible for energies close
to the double ionisation threshold and we are left with the $L=0$
classical Hamiltonian.	
			
	Finally, we point out that neither the Wannier orbit nor the Langmuir orbit
have appeared in our scaled spectroscopy.  According to 
reference~\cite{Muller92}, the later orbit should contribute
only for states above the $N=10$ threshold, which would explain that it is not
observed in our study. The fact that we find peaks only at the collinear orbits
can provide an explanation for the efficiency of the previous semiclassical
quantizations \cite{Richter90,Ezra91,Wintgen92}, which are thus legitimatized
by our scaled spectroscopy analysis.  Furthermore, our results support the
hypothesis that the semiclassical quantization of the ${}^1{\bf P}^o$ states
can in principle be achieved with these collinear orbits only.  Accordingly, the
non-zero angular momentum states could be semiclassically quantized starting
from these $L=0$ orbits.

\subsection{Wavefunctions}

	The similarity between the ${}^1{\bf P}^o$ and the 
${}^1{\bf S}^e$ states can be displayed by looking at
the wavefunctions. For the ${}^1{\bf S}^e$ states, there are only three
internal degrees of freedom, namely the three interparticle distances and there
is no angular dependency. This is no longer the case for the ${}^1{\bf P}^o$
states, in which the angular part is not constant (but given by the three
rotation matrix elements ${\cal D}^{1^{\displaystyle *}}_{0T}$). 
To allow pertinent comparison, we then trace the wavefunctions over the angular
variables and study the dependencies with respect to the three remaining degrees
of freedom. 

Since our scaled spectroscopy has shown the important role of collinear
configurations, we consider for our comparison the resonances satisfying the two
criterions~: (1) the  value of $\cos\theta_{12}$ has to be close to $1$
or $-1$; (2) the width is small compared to the distances to the
neighbouring states.  This last condition allows us to give a sense to the
corresponding  wavefunction because the resonance is long-lived and
isolated. Selecting wavefunctions according to the two above criteria, we find
that the resonances separate rather clearly into the two classes already
observed for the ${}^1{\bf S}^e$ states, namely the frozen-planet
configuration for which the two electrons are on the same side of the nucleus
and the asymmetric stretch configuration for which the two electrons are
localized on the opposite sides.

	Fig.~\ref{fro1po} depicts the lowest frozen-planet state in
the series $N=7$. Its energy is $-0.0113416$ atomic units (77.77 eV above the
helium ground state), its
width is $1.38\times 10^{-6}$ ($0.15$ meV). 
The real part of the  value of
$\cos\theta_{12}$ is  $0.73$, the imaginary part being hundred times smaller.
The figure shows the conditional density probability of the inner electron with
respect to the fixed axis ${\bf R}$ between the nucleus and the outer
electron. The distance  ${\bf R}$ is given by the classical expectation value
of the outer electron along the classical periodic orbit. Following the notation
of Ref.~\cite{Richter92}, this state is $(n,k,l)=(6,0,0)$. These integers
are the semiclassical quantum numbers in the EBK quantization of the
tori surrounding the periodic orbit. $n$ gives the number of nodes along the
orbit, while $k$ and $l$ are related the stable motion perpendicular 
to the orbit. More precisely, $k$ is the quantum number associated with the
bending degree of freedom and $l$ is associated with the stability of the 
orbit with respect to perturbation preserving the collinearity.
  In this figure, we see clearly the localization of the inner
electron on the same side of the nucleus. This figure compares very well with
the corresponding figures of Ref.~\cite{Richter92}. 

	Furthermore,  
Fig.~\ref{stretch1po} depicts a state in the $N=7$ series
($E=-0.011564$ in atomic units or 77.75 eV above the helium 
ground state) with a very small width 
($\Gamma=8.60\times 10^{-9}$ a.u. or 0.936 $\mu$eV). 
But, the  value of $\cos\theta_{12}$ is in this case equal to
$-0.842$. The electrons are essentially on the opposite side of the nucleus.
Thus, we have plotted the wavefunction in the $(r_1,r_2)$ plane for fixed
inter-electronic angle $\theta_{12}=\pi$. The maximum of probability is
clearly localized around the $-$ asymmetric stretch orbit, shown on the
top of the figure. Again this
figure is comparable to those obtained for ${\bf S}$ states, for example in
Ref.~\cite{Ezra91}.

\section{CONCLUSIONS}
\label{conclusion}

In this paper, we carried out an {\it ab initio} study of the $^1{\mathbf S}^e$
and $^1{\mathbf P}^o$ states of helium, for which we have
systematically calculated and characterized the main resonances and associated
wavefunctions.  Moreover, we have performed a semiclassical analysis by a
numerical scaled spectroscopy, which reveals the main periodic orbits emerging
out of the wave dynamics.  In order to probe the correlation between the two
electrons, we have studied, not only the mean level density and the
photoionization cross-section, but also the mean values of $\cos\theta_{12}$
where $\theta_{12}$ is the angle between the positions of both electrons with
respect to the nucleus.  The scaled spectroscopy based on this quantity
provides evidence for the collinear periodic orbis of both the $eZe$ and the
$Zee$ configurations.

Although the families of the $^1{\mathbf S}^e$ and $^1{\mathbf P}^o$ states are
ruled by different sets of partial differential equations derived from the
Schr\"odinger equation, the present study reveals a strong similarity between
these families, which can be explained by the difference of only one quantum of
angular momentum between them whereas they are highly excited in energy.  In
particular, periodic orbits emerge at comparable values of the scaled reduced
actions $S_p$ for both families.  this result is of great importance because it
shows that both families have similar semiclassical properties and that their
semiclassical quantization can in principle be carried out with 
collinear orbits of the same kind.  This conclusion is further supported by the
comparison between the resonant wavefunctions of the both families of states.

Moreover the calculated photoionization cross-section for the $^1{\mathbf P}^o$
states is in very good agreement with the recent experimental observations of
Refs.~\cite{Domke96}.  The observed overlap of the $N=5$ series with states of
the $N=6$ series is a quantum signature of the transition to a classically
chaotic regime, which is at the origin of a partial breakdown of Herrick's
classification scheme based on an integrable model.  When the overlapping of
the Rydberg series becomes important, Wigner repulsions between the states are
known to induce irregularities in the resonance spectrum.  In this regard, the
emerging periodic orbits revealed by the scaled spectroscopy provide a
complementary method to disclose ordered structures in highly congested spectra
like those of doubly-excited helium.

\vskip 1 cm

\noindent{\bf Acknowledgements}

It is our pleasure to thank Prof. G. Nicolis for support of this research.
P. G. is financially supported by the National Fund for Scientific Research
(F.N.R.S. Belgium).  During this research, B. G. has been financially supported
by a fellowship of the European Commission under contract No. ERBCHBICT941418. 
Part of this research has also been supported by the ARC project ``Quantum Keys
for Reactivity" of the ``Communaut\'e fran\c caise de Belgique" and by the
project ``Chaos and quantum mechanics in mesoscopic systems" of the ``Banque
Nationale de Belgique".

\begin{figure}
\caption{\label{fro1po} Conditional probability density for the inner
electron with respect to the fixed axis between the nucleus and the
outer electron. The outer electron is at a distance of $260$ atomic
units, the classical expectation value for the outer electron along
the classical $Zee$ periodic orbit. This resonance belongs to the
$N=7$ series of ${}^1{\bf P}^o$ states. 
Its energy is $-0.0113416$ atomic units (77.77 eV above the helium
ground state) and its width is
$1.38\times 10^{-6}$ (0.15 meV). The expectation value of $\cos\theta_{12}$ is $0.73$. The inner electron is clearly localized between the nucleus
(0,0) and the outer electron (260,0) and the wavefunction looks very much
like a hydrogenic Stark state, emphasizing the classical interpretation of
a inner electron oscillating in the electric field created by the outer one.} 
\end{figure}

\begin{figure}
\caption{\label{stretch1po} On the bottom~: 
probability density for the two electrons at
fixed inter-electronic angle $\theta_{12}=\pi$. This resonance is taken
from the $N=7$ series of ${}^1{\bf P}^o$ states. 
Its energy is $-0.011564$ in atomic units (77.75 eV above the helium
ground state) 
and its width
is $8.60\times 10^{-9}$ a.u. (0.936 $\mu$eV), which is very small compare to
other states in the same energy domain. The expectation value of
$\cos\theta_{12}$ is $-0.842$. It is strongly localized around the $-$
asymmetric stretch orbit (shown on the top of the figure), 
which is probably the reason of a so long lifetime.}
\end{figure}

\clearpage
\begin{table}
\caption{\label{po} Reduced actions $S_p(E=-1)/2\pi$ 
of several periodic orbits of helium.
Stability is with respect to deplacement conserving collinearity. F.P.
stands for frozen-planet, A.S. for asymmetric stretch and $eZe$ for the
other collinear orbits of this configuration .}
\begin{tabular}{clcc}  name & code
& reduced action $S_p/2\pi$ & Stability \\ \hline 
F.P. & & 0.7458 & stable \\ 
Langmuir & & 0.6761 & stable  \\ 
Wannier & & 1.7500 & inf. unstable \\ 
A.S. & $-$ & 0.9145 & unstable \\ 
$eZe$ & $+-$ & 1.8091 & unstable \\ 
$eZe$ & $++-$ & 2.6631 & unstable \\ 
$eZe$ & $+--$ & 2.6951 & unstable \\ 
\end{tabular}
\end{table}

\begin{table}
\caption{\label{data}Theoretical values of resonances energies $E$, linewidths
$\Gamma$, reduced linewidths $\Gamma^*$, reduced probabilities of 
transition $P^*$ and Fano $q$-parameters
for the two principal series below the $N=5$ threshold ($5,3_n$ and $5,1_n$).
$\Gamma$ and $\Gamma^*$ are in meV.
The reduced probabilities values are given in percentage relatively to the
$2,0_n$ series. There are sensible changes of the properties for the states with
greater energy than $76.5915$ eV, the energy of the  interfering state
$6,4_6$.} \begin{tabular}{rccccc} 
$n$ & $E-E_g$ (eV) & $\Gamma$ & $\Gamma^*$ & $P^*$ & $q$  \\
\hline
\multicolumn{6}{c}{$5,3_n$} \\
 5 & 75.5639 & 59.09 & 2090 & 1.767 & -0.225 \\
 6 & 76.0840 & 42.49 & 3332 & 3.427 & -0.022 \\
 7 & 76.3199 & 25.16 & 3501 & 3.747 & -0.006 \\
 8 & 76.4605 & 12.41 & 2813 & 2.987 &  0.006 \\
 9 & 76.5479 &  3.02 & 1031 & 1.005 &  0.066 \\
10 & 76.5953 &  3.51 & 1588 & 1.043 &  0.537 \\
11 & 76.6278 & 11.89 & 6738 & 8.167 &  0.190 \\
12 & 76.6633 &  7.88 & 5998 & 7.596 &  0.046 \\
13 & 76 6926 &  4.97 & 5097 & 6.542 &  0.012 \\
14 & 76.7151 &  3.28 & 4438 & 5.870 &  0.024 \\
\multicolumn{6}{c}{$5,1_n$} \\
 5 & 75.7590 & 88.82 & 4041 & 0.176 &  0.437 \\
 6 & 76.2237 & 57.25 & 6140 & 0.170 &  0.706 \\
 7 & 76.4131 & 38.10 & 7196 & 0.222 &  0.702 \\
 8 & 76.5234 & 22.48 & 6762 & 0.286 &  0.629 \\
 9 & 76.5941 & 11.16 & 5000 & 0.707 &  0.023 \\
10 & 76.6422 &  4.35 & 2762 & 0.002 &  0.990 \\
11 & 76.6754 &  0.64 &  550 & 0.006 &  0.737 \\
12 & 76.6980 &  0.41 &  447 & 0.002 & -1.880 \\
\end{tabular}
\end{table}


\begin{thebibliography}{99}
\bibitem[(a)]{LKB} Present Address~: Laboratoire Kastler Brossel \\
Universit\'e Pierre et Marie Curie Tour 12 Etage 1 \\
4, place Jussieu \\
F-75252 Paris cedex 05


\bibitem{Richter90} Richter K. and Wintgen D., J. Phys. B \textbf{23}, L197
(1990)
\bibitem{Richter902} Richter K. and Wintgen D. 
J. Phys. B: At. Mol. Opt. Phys. \textbf{23} L197 (1990)
\bibitem{Ezra91} Ezra G.S., Richter K., Tanner G. and Wintgen D., J. Phys. B
\textbf{24}, L413 (1991)
\bibitem{Blumel91} Bl\"umel R. and Reinhard W.P. Direction in Chaos
\textbf{4} (1991)
\bibitem{Richter91} Richter K. and Wintgen D.,
J. Phys. B: At. Mol. Opt. Phys. \textbf{24} L565 (1991)
\bibitem{Richter92} Richter K., Briggs J.S., Wintgen D. and  Solov'ev E.A.,
J. Phys. B: At. Mol. Opt. Phys. \textbf{25} 3929 (1992)
\bibitem{Gaspard93} Gaspard P. and Rice S.A. Phys. Rev. A \textbf{93}
54 (1993)
\bibitem{Tanner95} Tanner G. and Wintgen D. Phys. Rev. Lett. \textbf{75}
2928 (1995)
\bibitem{Burgers95} B\"urgers A., Wintgen D., and Rost J.M. 
J. Phys. B: At. Mol. Opt. Phys. \textbf{28} 3163 (1995)
\bibitem{Gutzwiller90} \textit{Chaos in Classical and Quantum Mechanics}
Gutzwiller M.C. Springer-Verlag (1990)
\bibitem{Domke91} Domke M., Xue C., Puschmann A., Mandel T., Hudson E.
Shirley D.A., Kaindl G., Greene C.H., Sadeghpour H.R. and Petersen H.  
Phys. Rev. Lett. \textbf{66} 1306 (1991)
\bibitem{Domke92} Domke M., Remmers G., and Kaindl G. Phys. Rev. Lett.
\textbf{69} 1171 (1992)
\bibitem{Domke95} Domke M., Schulz K., Remmers G., Kaindl G., and Wintgen D.,
Phys. Rev. A, \textbf{53}, 1424 (1996)
\bibitem{Domke96}Schultz K.,  Kaindl G. Domke M. Bozek J.D. Heimann P.A.
Schlachter A.S. and Rost J.M. Phys. Rev. Lett. 
\textbf{77} 3086 (1996)
\bibitem{Wintgen92} Wintgen D., Richter K., and  Tanner G., Chaos \textbf{2}, 19
(1992)
\bibitem{Muller92} M\"uller J., Burgd\"orfer J., and Noid D., Phys. Rev. A
\textbf{45}, 1471 (1992)





\bibitem{Madden63} Madden R.P. and Codling K. Phys. Rev. Lett.
\textbf{10} 516 (1963)

\bibitem{Sinanoglu73} Sinanoglu O. Herrick D. R. J. Chem. Phys.
\textbf{62} 886 (1973)

\bibitem{Herrick75} Herrick D.R. Phys. Rev. A \textbf{12} 413 (1975)

\bibitem{Sinanoglu75} Herrick D.R. and Sinanoglu O. Phys. Rev. A
\textbf{11} 97 (1975)

\bibitem{Herrick77} Herrick D.R. Phys. Rev. A \textbf{17} 1 (1978)

\bibitem{Herrick80} Herrick D.R. and Poliak R.D. 
J. Phys. B: At. Mol. Opt. Phys. \textbf{13} 4533 (1980)

\bibitem{Herrick83} Herrick D.R. Adv. Chem. Phys. \textbf{52} 1 (1983)

\bibitem{Feagin86} Feagin J.M. and Briggs J.S. Phys. Rev. Lett.
\textbf{57} 984 (1986)

\bibitem{Feagin88} Feagin J.M. and Briggs J.S. Phys. Rev. A
\textbf{37} 4599 (1988)

\bibitem{Rost88} Rost J.M. and Briggs J.S. 
J. Phys. B: At. Mol. Opt. Phys. \textbf{21} L233 (1988)

\bibitem{Kellman94} Kellman M.E. Phys. Rev. Lett. \textbf{73} 2543
(1994)


\bibitem{Ho86} Ho Y.K., Phys. Rev. A, \textbf{34} 4402 (1986)

\bibitem{Watanabe86} Watanabe S. and Lin C.D. Phys. Rev. A 
\textbf{34} 823 (1986)
\bibitem{Dmitrieva86} Dmitrieva I.K. and Plindov G.I. J. Physique
\textbf{47} 1493 (1986)
\bibitem{Rost91} Rost J.M., \textit{et al}. 
J. Phys. B: At. Mol. Opt. Phys. \textbf{24} 2455 (1991)

\bibitem{Briggs91} Rost J.M. and Briggs J.S.
J. Phys. B: At. Mol. Opt. Phys. \textbf{24} 4293 (1991)


\bibitem{Ho93} Ho Y.K., Phys. Rev. A, \textbf{48} 3598 (1993)
\bibitem{Ostrovsky93} Ostrovsky V.N. and Prudov N.V. 
J. Phys. B: At. Mol. Opt. Phys. \textbf{26} L263 (1993)
\bibitem{Muller94} M\"uller J., Yang X., and Burgd\"orfer J., 
Phys. Rev. A \textbf{49} 2470 (1994)






\bibitem{perimetric} Gr\'emaud B. and Delande D. (to be published)


\bibitem{Ho83} Ho Y.K., Phys. Rep. \textbf{99} 1-68 (1983)

\bibitem{Balslev71} Balslev E. and Combes J.M., Commun. Math. Phys.
\textbf{22} 280 (1971)


\bibitem{Buchl94} Buchleitner A., Gr\'emaud B. and Delande D.,
J. Phys. B: At. Mol. Opt. Phys. \textbf{27} 2663 (1994)

\bibitem{Burgers96} B\"urgers A. and Rost J.M.,
J. Phys. B: At. Mol. Opt. Phys. \textbf{29} 3825 (1996)







\bibitem{Richter93} Richter K., Tanner G. and Wintgen D. 
Phys. Rev. A \textbf{48} 4182 (1993)


\bibitem{Gaspard95} Gaspard P., Alonso D., and Burghardt I. 
Adv. Chem. Phys. \textbf{XC} 105 (1995) 



\bibitem{section} Gr\'emaud B. and Delande D.(to be published)








\end{thebibliography}
\end{document}